\documentclass[amsmath,amssymb,pra,twocolumn,superscriptaddress]{revtex4-1}

\usepackage{amsmath}
\usepackage{amsfonts}
\usepackage{graphicx}
\usepackage{dcolumn}
\usepackage{amssymb}
\usepackage{physics}
\usepackage[separate-uncertainty = true]{siunitx}

\usepackage{hyperref}
\usepackage{subfig}
\usepackage{mhchem}
\usepackage{bm}
\usepackage{pgfplots,relsize}
\usepackage{tikz}

\usepackage{chemformula}
\usepackage{chemfig}      
\usepackage{siunitx}
\usepackage{caption}
\usepackage{comment}

\newcommand{\RN}[1]{%
  \textup{\uppercase\expandafter{\romannumeral#1}}%
}

\begin{document}

\title{Doubly resonant photonic crystal cavities for efficient\\ second-harmonic generation in III-V semiconductors}

\author{Simone Zanotti}
\affiliation{Dipartimento di Fisica, Universit\`a di Pavia, via Bassi 6, 27100 Pavia (Italy) }

\author{Momchil Minkov}
\affiliation{Department of Electrical Engineering, and Ginzton Laboratory, Stanford University, Stanford, California 94305 (USA) }

\author{Shanhui Fan}
\affiliation{Department of Electrical Engineering, and Ginzton Laboratory, Stanford University, Stanford, California 94305 (USA) }

\author{Lucio C. Andreani}
\affiliation{Dipartimento di Fisica, Universit\`a di Pavia, via Bassi 6, 27100 Pavia (Italy) }
\affiliation{Institute for Photonics and Nanotechnologies (IFN)-CNR, 20133 Milano (Italy) }

\author{Dario Gerace}\email{dario.gerace@unipv.it }
\affiliation{Dipartimento di Fisica, Universit\`a di Pavia, via Bassi 6, 27100 Pavia (Italy) }

\begin{abstract}
Second-order nonlinear effects, such as second-harmonic generation, can be strongly enhanced in nanofabricated photonic materials when both fundamental and harmonic frequencies are spatially and temporally confined. Practically designing low-volume and doubly resonant nanoresonators in conventional semiconductor compounds is challenging owing to their intrinsic refractive index dispersion. In this work we review a recently developed strategy to design doubly resonant nanocavities with low mode volume and large quality factor by localized defects in a photonic crystal structure. We build on this approach by applying an evolutionary optimisation algorithm in connection with Maxwell equations solvers, showing that the proposed design recipe can be applied to any material platform. We explicitly calculate the second-harmonic generation efficiency for doubly resonant photonic crystal cavity designs in typical III-V semiconductor materials, such as GaN and AlGaAs, targeting a fundamental harmonic at telecom wavelengths, and fully accounting for the tensor nature of the respective nonlinear susceptibilities. These results may stimulate the realisation of small footprint photonic nanostructures in leading semiconductor material platforms to achieve unprecedented nonlinear efficiencies. 
\end{abstract}

\maketitle

\section{Introduction}
\begin{figure*}
\centering
\includegraphics[scale=0.7]{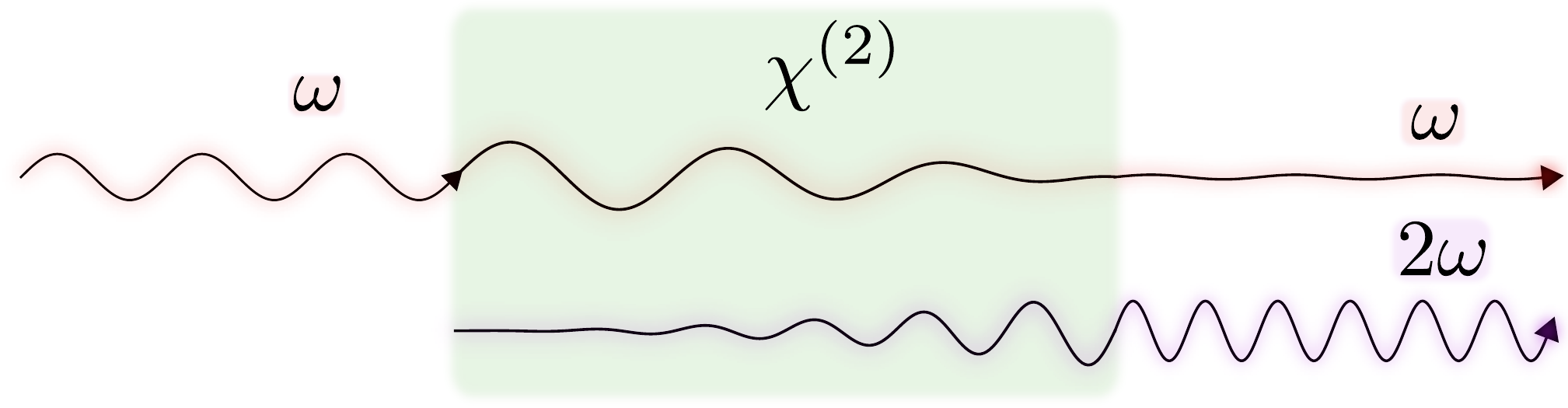}
\caption{Schematic representation of the second-harmonic generation process, where the electromagnetic radiation is converted from the frequency $\omega$ to its double frequency $2\omega$.}\label{fig:SHG}
\end{figure*}

Nonlinear optical processes are notoriously poorly efficient, due to small higher-order nonlinear susceptibilities $(\chi^{(n)})$ that mediate such processes in conventional materials such as semiconductors or insulators in their transparency spectral range \cite{book:boyd}. A way out this limiting factor can be tackled by strongly confining the electromagnetic radiation in dielectric resonators made of such nonlinear materials. The confinement of the radiation in a small volume ($V\sim\lambda^3$) and for a long time (high-quality factor $Q$) allows for a strong interaction between the electromagnetic field and the nonlinear dielectric, thus enhancing the efficiency of nonlinear effects \cite{art:rodr}.
Among the possible nanoresonators, photonic crystal (PhC) cavities offer one of the most promising platforms to achieve such extreme confinement conditions. In the last few years, many efforts have been made to achieve enhanced second-order nonlinearities in both singly-resonant PhC cavities \cite{art:singres,art:sonia}, where the radiation is trapped only for the first-harmonic (FH) frequency, $\omega_{\mathrm{FH}}=\omega$, and doubly-resonant PhC cavities \cite{art:rodr,art:doub1D,art:momch}, in which the confinement simultaneously occurs for both FH and second-harmonic (SH) frequency, $\omega_{\mathrm{SH}}=2\omega$.
Hence, doubly-resonant PhC cavities are also among the most promising choices for enhancing second-order nonlinear effects, e.g. second harmonic-generation (SHG) (see Figure \ref{fig:SHG}) and spontaneous parametric down-conversion (SPDC), since their efficiency benefits from the simultaneous confinement of both FH and SH modes, provided they are coupled by the nonlinear tensor elements.
While PhC cavities possess several degrees of freedom to be used in order to tailor their optical properties, very few of such devices have been studied to fulfil a doubly resonant condition that could boost SH generation efficiency, until very recently \cite{art:momexp}. The main obstacle comes from the difficulty in designing PhCs with photonic band gaps around the FH and SH frequencies, which would allow obtaining two doubly-resonant cavity modes \cite{art:triang}. Recently, a novel approach based on the so-called bound-states in the continuum (BIC) has been proposed \cite{art:momch}, overcoming the aforementioned obstacle. This strategy has later been shown to be practically effective by realizing doubly resonant SH generation in a III-V semiconductor PhC cavity \cite{art:momexp}. The BICs are particular states that stay confined despite the fact they lie above the light line, i.e., their dispersion falls within the continuum of radiative modes \cite{rev:BIC}. This feature is indispensable to achieve good temporal confinement of the SH cavity mode, even in the absence of a photonic bandgap \cite{art:momch}.

In the present article, we review the design strategy based on matching the FH of a PhC lattice with a suitable BIC mode at SH frequency, and we generalize the procedure to show that doubly-resonant PhC cavities based on BICs can be properly designed by applying a particle swarm optimisation (PSO) algorithm combined with a numerical solver for Maxwell equations \cite{art:PSO}; in particular, we use the finite-difference time-domain (FDTD) algorithm here \cite{book:love}, which allows a precise determination of the higher order mode resonances and is amenable to be extended to large cavity structures. The latter are shown to allow for both FH and SH mode confinement in the plane through a heterostructure engineering of the PhC lattice. Moreover, in determining the nonlinear properties we directly take into account the tensor nature of the second-order nonlinear susceptibility, $\chi_{i,j,k}^{(2)}$ \cite{art:rodr}, a key feature to consider when estimating their efficiency, which is sometimes overlooked in the literature. This allowed us to numerically calculate the SHG efficiency and its dependence on the crystal growth direction with respect to the PhC cavity axes, through a dimensionless overlap integral ($\bar{\beta}$) between the FH and SH modes taking into account the $\chi^{(2)}$ tensor. The latter can be regarded as a generalisation of the phase-matching condition for the confined systems. The factor $\bar{\beta}$ is directly related to the estimation of the SHG efficiency, which can be approximated to scale as $Q_1^2Q_2|\bar{\beta}|^2$, $Q_1$ and $Q_2$ being the FH and SH modes quality factors, respectively \cite{art:rodr,art:doub1D}. Our numerical calculations show that different crystal structures lead to different dependencies of the SHG efficiency on the crystal axes orientation. In particular, we investigated the two most common crystal structures in $\RN{3}\!-\!\RN{5}$ semiconductors, namely the zincblende and the wurtzite whose typical examples are \ch{AlGaAs} and \ch{GaN}, respectively. These compounds have gained increasing interest in nonlinear applications due to their strong nonlinear properties, and for being transparent at both FH/SH frequencies targeted at telecom/near infrared wavelengths, i.e., $\lambda_{\text{FH}}\approx$ \SI{1550}{\nano\meter} and $\lambda_{\text{SH}} \approx$ \SI{775}{\nano\meter}. 
It is found that the $[0,0,1]$ growth direction of the zincblende crystalline materials lead to a vanishing overlap integral for specific orientations of the crystal axes with respect to the PhC lattice. In these unfavourable cases, the conversion efficiency is theoretically suppressed, despite the strong temporal and spatial confinements. However, we show in the present work that a suitable growth direction or crystal structure, such as $[1,1,1]$ grown zincblende, or the wurtzite structure, lead to an isotropic generation efficiency that does not depend on the PhC orientation against the crystal unitary cell.

The manuscript is organized as follows. In Sec.~\ref{sec:phc} we review the design procedure to obtain a doubly resonant photonic crystal structure in which the SH mode is a BIC, generalize it by combining Maxwell solvers with evolutionary optimization, and explicitly apply it to the most promising III-V platforms in nonlinear optics. In Sec.~\ref{sec:doub_cav} we show that a simultaneous spatial confinement of both FH and SH modes can be achieved by applying a varying hole radii photonic heterostructure. In Sec.~\ref{sec:overlap} we calculate the nonlinear conversion efficiency in such doubly resonant photonic crystal cavities, taking into account the crystal structure of the III-V material and the orientation of the photonic lattice with respect to the semiconductor unit cell. 

\section{Evolutionary optimization of doubly resonant photonic crystals}\label{sec:phc}
\subsection{Numerical methods}\label{sec:meth}
The present design strategy to obtain doubly-resonant PhC cavities relies on two numerical methods to solve Maxwell equations in nanofabricated dielectric materials, namely the guided-mode expansion (GME) and the FDTD methods. The former method is well suited to the analysis of layered systems with in-plane periodicity of the dielectric function, such as PhC slabs, providing a solution for the frequency-wavevector dispersion of the real and imaginary part of complex eigenmodes \cite{art:triang}. Here, we consider only PhC slabs with holes carved in the dielectric slab with a triangular lattice, but these are not intrinsic constraints and other lattices could be considered as well. In this case, the main parameters are the lattice constant $a$, and the (dimensionless) radius $r/a$ and slab thickness $d/a$. The GME starts from a decomposition of the electromagnetic field in a basis of plane waves and guided modes of an effective planar waveguide, which leads to an eigenvalue problem whose eigenvalues correspond to the real part of the complex eigenfrequencies of the PhC modes \cite{art:triang}. Moreover, perturbation theory allows to calculate the coupling of the PhC slab modes to the out-of-plane propagating radiation modes of the effective planar waveguide with average refractive index, which can be considered the loss rate of the PhC slab modes defined as the imaginary part of the complex eigenmode frequencies \cite{art:nearlyfree}. This approach turns out to be computationally very efficient and quantitatively accurate to within a few percent error \cite{art:momdiff}, albeit approximate and polarization-dependent accuracy. Hence, it has been applied for a rapid and coarse scanning of the parameters, and for a fast identification of higher order modes in the SH frequency range. 

In parallel, the FDTD method provides the electrodynamics solution of Maxwell equations for arbitrary complex systems, by discretizing and solving the finite difference formulation of the coupled differential equations \cite{book:love}. The FDTD algorithm provides a numerically exact method, i.e. with sufficiently fine discretization its solution should converge to the exact results. In this case, the imaginary part of the frequency of a given electromagnetic mode of a structure is obtained by fitting the exponential decay of the fields associated to that mode. This is best achieved by selectively exciting the mode using symmetry considerations, and waiting for low-Q modes to decay first. Since the FDTD is accurate but computationally costly, it has been applied only after the target modes have been identified with the GME solution. Throughout this work, we have employed a commercial implementation of the three dimensional (3D) FDTD solver provided from Lumerical-Ansys. Moreover, a particle swarm optimisation (PSO) algorithm has been exploited in connection with the 3D-FDTD solver in order to finely tune the PhC parameters \cite{art:PSO}. Such a routine allows to efficiently minimise a given function $f:A\to \mathbb{R}$, also named figure of merit, by varying the parameters in the search-space $A\subset\mathbb{R}^n$, further information on this algorithm are provided in Appendix \ref{sec:PSO}. In the present case, the PSO was chosen since it allows to reach convergence within a few evolutionary steps when the starting point is not too far from the target objectives, as it is most often the case when targeting the doubly resonant condition.

\subsection{Design procedure}
The present design strategy to obtain doubly-resonant PhC cavities builds on a previously presented band engineering recipe, aimed at finding a SH band with a mode displaying infinite lifetime for emission along the vertical direction with respect to the PhC plane, i.e. a BIC \cite{art:momch}. This approach has also been recently shown to experimentally  enhance SHG efficiency in GaN PhC cavities \cite{art:momexp}. Here we show that the above strategy can be further generalized and automated, and it can be applied to an arbitrary material platform. The procedure can be summarised as follows:
\begin{enumerate}
\item	Calculate the photonic bandstructure of the PhC slab with GME.
\item	Identify a band-edge mode below the light line as FH mode at frequency $\omega_1$, and a non-degenerate BIC mode above the light line at frequency $\omega_2$ as SH mode, whose photonic mode dispersion depends on the refractive indices of the target material.
\item Check that FH and SH modes have non-vanishing overlap integral $\bar{\beta}$, depending on the $\chi^{(2)}$ tensor of the given material.
\item Vary the hole radius and the slab thickness in order to approximately match the doubly resonant condition $\omega_2\approx2\,\omega_1$.
\item Starting from the parameters calculated with GME, apply the PSO in connection with the FDTD Maxwell solver to finely tune the doubly resonant condition and reach $\omega_2=2\,\omega_1$ within the accepted tolerance.
\item Spatially confine the FH and SH mode by suitably designing a heterostructure PhC cavity.
\end{enumerate}

\subsection{Doubly-resonant photonic crystal slabs}
\begin{figure*}
\centering
\includegraphics[width=\linewidth]{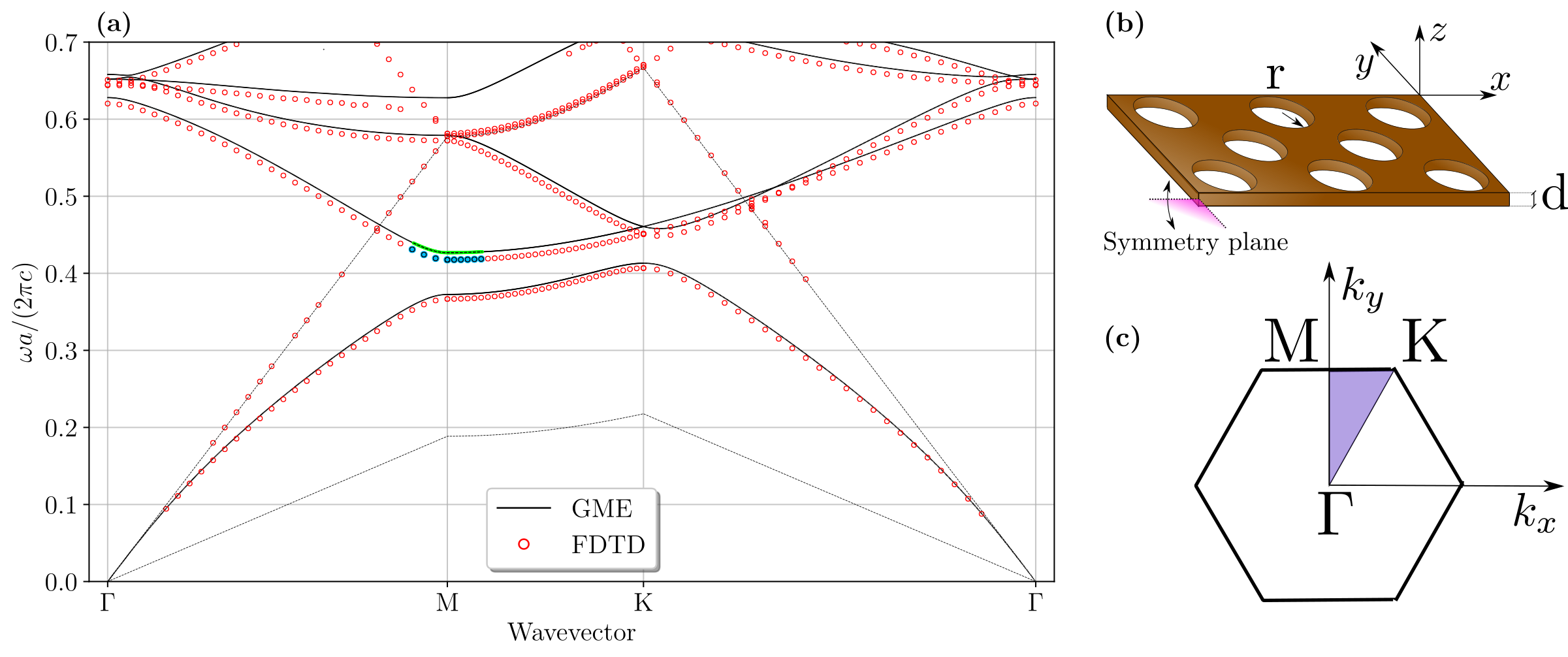}
\caption{\textbf{(a)} Real part of the \ch{GaN} even modes (TE-like)  bandstructure calculated with the GME and the FDTD methods, respectively. The parameters chosen for these simulations are: $n=2.28$, $d/a=0.337$ and $r/a=0.206$. The $k-$space region surrounding of the FH mode at the M point is highlighted, while the light lines are dashed. \textbf{(b)} Representation of the thin dielectric slab with a triangular photonic lattice, with an indication of the relevant symmetry plane (at $z=0$). \textbf{(c)} Main symmetry points considered within the Brillouin zone.}\label{fig:TE}
\end{figure*}  

\begin{figure*}
\centering
\includegraphics[width=\linewidth]{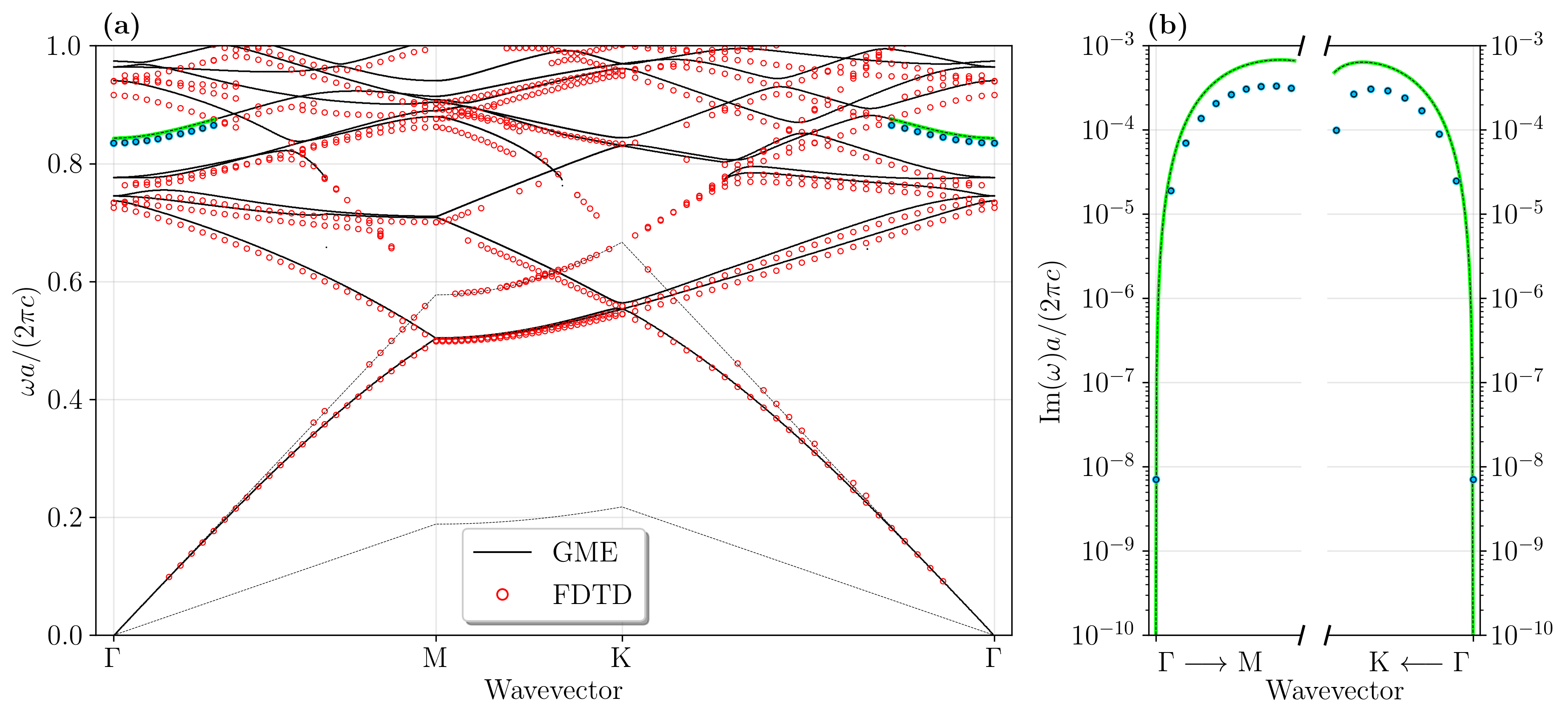}
\caption{\textbf{(a)} Real part of the \ch{GaN} odd modes (TM-like) bandstructure calculated with the GME and the FDTD methods. The parameters chosen for these simulations are: $n=2.31$, $d/a=0.337$ and $r/a=0.206$. The $k-$space region surrounding the SH mode at the $\Gamma$ point is highlighted. \textbf{(b)} Imaginary parts of the eigenmodes highlighted in the bandstructure \textbf{(a)}. The vanishing Im$(\omega)$ indicates the presence of a BIC at the $\Gamma$ point, both from GME and FDTD solutions.}\label{fig:TM}
\end{figure*}

With the target of optimizing the doubly resonant condition for III-V semiconductors as the relevant nonlinear materials, we consider that FH mode with dominant in-plane polarization might be coupled by the second-order tensor elements to SH modes with dominant vertical polarization. For this reason, we consider the photonic mode dispersion calculated for membrane PhC slabs (i.e., with symmetric air claddings above and below the patterned semiconductor), with FH modes having even symmetry (transverse-electric, TE-like) with respect to the horizontal ($x,y$) plane bisecting the planar waveguide, and SH modes possessing odd symmetry (transverse magnetic, TM-like) \cite{art:triang,art:momdiff}. These are shown for the specific case of a \ch{GaN} PhC slab in Figs. \ref{fig:TE} and \ref{fig:TM}, respectively. The photonic lattice with the definition of the relevant real space axes and the Brillouin zone in reciprocal space are also shown, in Fig.~\ref{fig:TE}(b)-(c). In order to take into account the material dispersion, the two solutions have been obtained with different refractive indices: for the FH modes, we set $n=2.28$, while for SH modes $n=2.31$. These values correspond to the \ch{GaN} refractive indices at the target wavelengths $\lambda_{\text{FH}}=$ \SI{1550}{\nano\meter} and $\lambda_{\text{SH}} =$ \SI{775}{\nano\meter}, respectively. The numerical solutions are explicitly shown for both GME and 3D-FDTD solvers, and directly compared to each other. For the FDTD solution, broad-band, point-like electric dipole sources are randomly placed in the whole elementary cell in order to excite all the possible modes, and proper phases are given to each dipole in order to satisfy the Bloch boundary conditions for given in-plane wave vectors; thus, each wave vector in the first Brillouin zone of the corresponding photonic lattice requires a separate FDTD simulation, and by running over all the k-vectors, the whole band structure is reconstructed. \\
The dispersion of photonic bands (real part of complex eigenfrequencies) is shown for both FH and SH modes in Figs.~\ref{fig:TE}(a) and \ref{fig:TM}(a), while the imaginary part is directly compared only for the target BIC mode in the SH case, Fig.~\ref{fig:TM}(b). It is worth stressing that the results obtained with the two methods are in overall good agreement, quite remarkably considering that they rely on conceptually different methods, and moreover the GME method is an approximate method. These results suggest that the GME method is quite effective in simulating such photonic crystal devices. Notice that the imaginary part of the BIC mode goes to zero at the $\Gamma$ point in both solutions. The agreement is slightly worse for the odd modes, which may also expected owing to the approximations inherent to the GME approach. Hence, this comparison justifies employing GME for a fast parameter scan, while the FDTD method is particularly suited to give accurate results in the SH case with TM-like symmetry.

In view of defining confined modes, we identify the FH mode with the second band of the TE-like bandstructure at the M point, which is located below the air light line and is likely to give rise to a high quality factor cavity mode upon confinement \cite{art:momch}. On the other hand, the SH mode inevitably lies above the air light line, as it is clear from the TM-like bandstructure; for this reason, one of the most promising modes to consider for a potentially high quality factor confined mode at SH is the seventh band at $\Gamma$, which is  a BIC in the $\Gamma$ point as already stressed. These modes are selected as a starting point for the evolutionary optimization aimed at searching the photonic structure parameters giving doubly resonant condition. 

Once the candidate FH and SH modes are identified with the GME solution, a coarse span of the parameters is necessary to "move close" to the doubly resonant condition. At this point, the PSO can be implemented in the FDTD algorithm to accurately tune the PhC slab parameters. In this case, we let the optimisation algorithm vary the radius of the holes and the slab thickness in the search-space $A =[r_\text{min},r_\text{max}]\times[d_\text{min},d_\text{max}]$, where $A$ is a small neighbourhood around the best GME parameters; while the figure of merit to be minimised has been defined as:
\begin{equation}
    \text{FOM}=\bigg\lvert\frac{\omega_2-2\omega_1}{\omega_2}\bigg\rvert,
\end{equation}
where $\omega_1$ and $\omega_2$ are the FH and SH mode frequency, respectively. We imposed the condition $\text{FOM}\leq 1\%$ to be fulfilled, such that the doubly resonant condition is achieved with a given tolerance.

\begin{table*}
\caption{Main design parameters of the \ch{GaN} and \ch{Al_{0.3}Ga_{0.7}As} PhC slabs obtained with the PSO algorithm applied to a 3D-FDTD solver.}\label{tab:res}
\centering
\begin{tabular}{cccccccccc}
\toprule
\textbf{Material} &$n_{\text{FH}}$ &$n_{\text{SH}}$	& $d/a$	& $r/a$ &$\omega_1 a/(2\pi c)$ & $\omega_2 a/(2\pi c)$ & $a$ (\si{\nano\meter})& $d$ (\si{\nano\meter})	& $r$ (\si{\nano\meter})    \\
\ch{GaN}		& $2.28$ & $2.31$  &$0.337$ 		& $ 0.206$ & $0.414$& $0.832$& $642.3$	& $216.4$			& $132.2$ \\
\ch{Al_{0.3}Ga_{0.7}As}	& $3.23$ & $3.47$	& $0.24$			& $0.175$ &$0.3356$ & $0.6650$& $520.1$	& $124.8$			& $91.2$\\
\toprule
\end{tabular}
\end{table*}

To show the generality of the approach, the very same procedure has been applied to both \ch{GaN} and \ch{Al_x Ga_{1-x} As} materials (assuming $x=0.3$ for the choice of room temperature refractive indices), targeting  the same bands in the same high symmetry points within the Brillouin zone to optimize the doubly resonant condition. All the quantities are summarised in dimensionless units in table \ref{tab:res}. They can also easily be converted to dimensional units by imposing the target wavelength, e.g. $\lambda_\text{FH}=$ \SI{1550}{\nano\meter} for the FH mode, yielding  $a=\tilde{\omega}_1\lambda_{\text{FH}}$, $\tilde{\omega}_1$ being the FH dimensionless frequency, from which the lattice constant $a$, and all other quantities, can be calculated.

\begin{figure}[b]
\centering
\includegraphics[width=\linewidth]{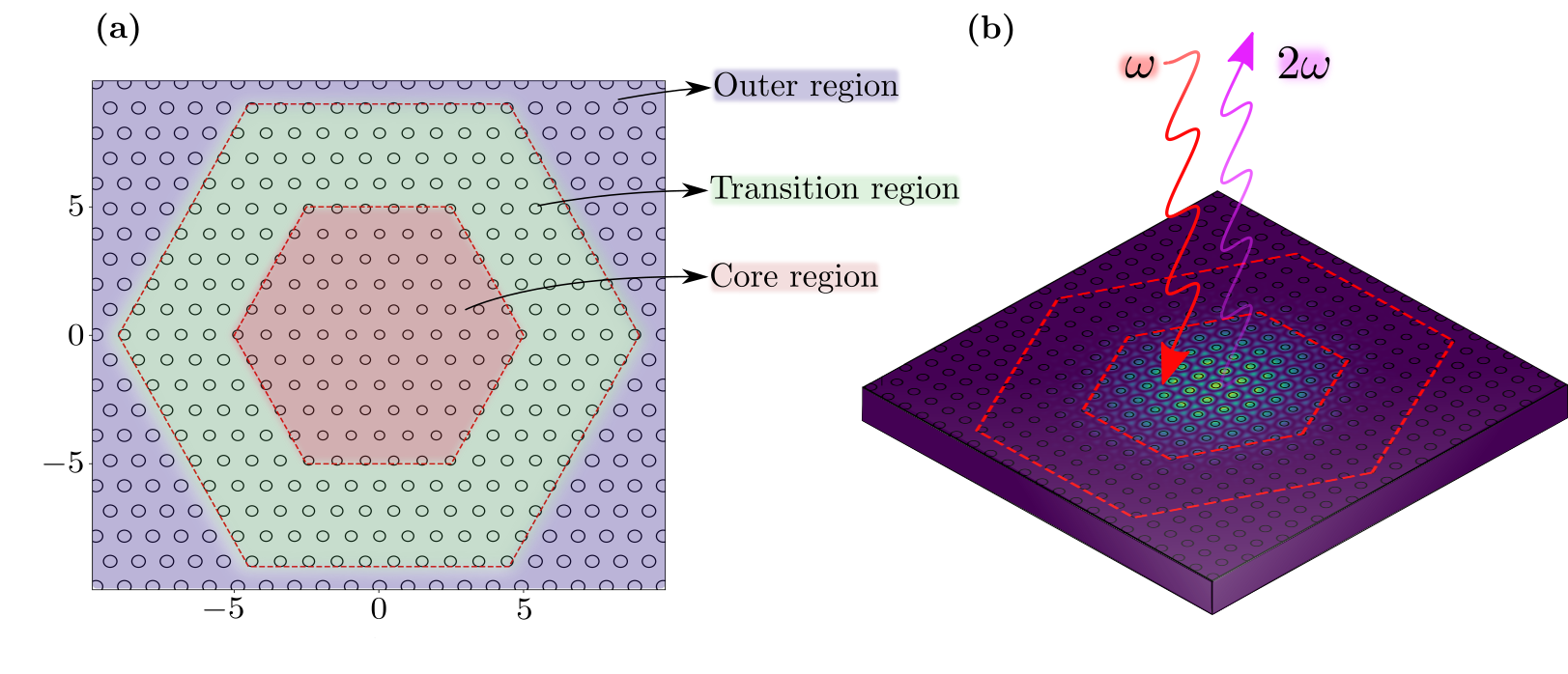}
\caption{\textbf{(a)} The heterostructure cavity is defined by core, transition, and outer regions. The hole radius is gradually increased on going outwards from the core region. \textbf{(b)} Schematic representation of the SHG process in the doubly resonant PhC cavity.}\label{fig:cav_3D}
\end{figure}  

\section{Doubly-resonant photonic crystal cavities }\label{sec:doub_cav}
In the previous section we have outlined how to achieve the doubly resonant condition in PhC slabs. While in this situation the electromagnetic field is bound in the vertical direction, $z$, it is still fully delocalized in the $x,y$ plane. In view of enhancing the nonlinear efficiency, it is necessary to introduce a heterostructure design to achieve the field confinement also along the other two spatial directions. Such a heterostructure ultimately aims at producing a confining potential for the photons, under the very same principle allowing for charged carriers to be confined within a semiconductor heterostructure. As it is commonly observed,  for the photonic crystal slab system considered here, the resonators with highest quality factors tend to confine photons in the regions of higher dielectric constant \cite{art:cav_mom_orig}. Hence, the cavity is designed by assuming three concentric hexagonal regions with gradually increasing radii, such that an effective potential well is created for photons. The structure geometry is shown in Fig.~\ref{fig:cav_3D}, and it is characterized by the side length of the hexagons ($N_c,N_t,N_o$) in terms of lattice constant and the holes radii ($r_c,r_t,r_o$), in which the subscripts ($c,t,o$) refer to core, transition and outer regions, similarly to Ref.~\cite{art:momch}. For consistency, we set $N_c=6$, $N_t=4$ and $N_o=10$, for simulations on both \ch{GaN} and \ch{Al_{0.3}Ga_{0.7}As} platforms.
The slab thickness and the hole radius in the core region are inherited from the PhC slabs parameters, while the radii of holes within the transition and outer regions are slightly increased to allow for a gentle confinement of the electromagnetic field. Due to small frequency shifts, a slight tuning of the parameters may be necessary in order to restore the doubly resonance condition in the cavity design.

\begin{table*}
\caption{Main design parameters of the \ch{GaN} and \ch{Al_{0.3}Ga_{0.7}As} cavities, and the related doubly resonant features obtained from 3D FDTD simulations.}\label{tab:cav}
\centering
\begin{tabular}{cccccccccc}
\toprule
\textbf{Material}  &$a$  (\si{\nano\meter})& $d$  (\si{\nano\meter}) & $r_c$ (\si{\nano\meter})& $r_t$ (\si{\nano\meter})	& $r_o$ (\si{\nano\meter})&$\lambda_{\text{FH}}$ &$\lambda_{\text{SH}}$	& $Q_{\text{FH}}$	& $Q_{\text{SH}}$    \\
\ch{GaN}	& $642$& $200$& $122$	& $145$			& $165$	& $1503$ & $753$  &$3.6\cdot10^4$ 		& $1100$ \\
\ch{Al_{0.3}Ga_{0.7}As}	 &$520$ & $124$& $91$	& $105$ & $121$ & $1523$ & $759$ 	&	$1.1\cdot10^5$		& $400$\\
\toprule
\end{tabular}
\end{table*}

\begin{figure}[b]
\centering
\includegraphics[width=\linewidth]{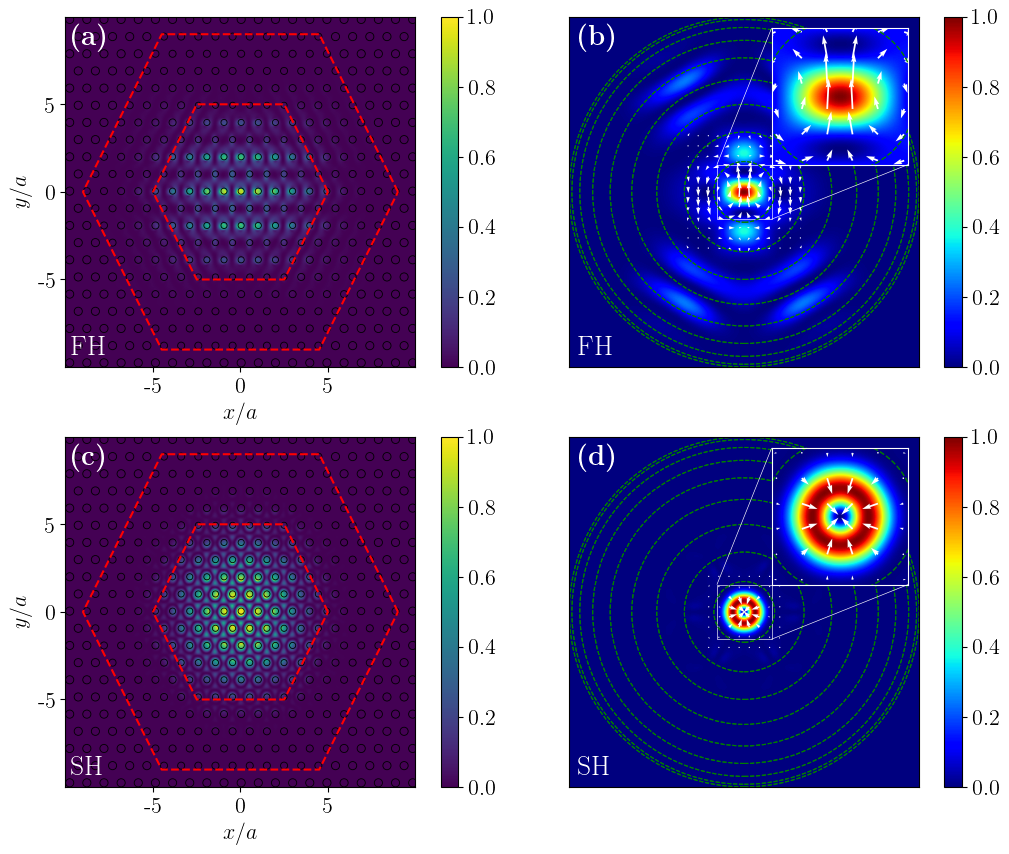}
\caption{\textbf{(a)}-\textbf{(c)} FH and SH cavity mode intensity profiles at the centre of the slab $z=0$. \textbf{(b)}-\textbf{(d)} Colour maps representing the emission intensity profile of the FH and SH cavity modes. Every green circle identifies \SI{10}{\degree} from the vertical direction of emission with respect the slab plane. The white arrows show the vector field defined by ($\Re{\left(E_x\right)},\Re{\left(E_y\right)}$).}\label{fig:cav_both}
\end{figure} 

We studied the properties of the same heterostructure cavity design realized in both \ch{GaN} and \ch{Al_{0.3}Ga_{0.7}As} materials, by using 3D FDTD simulations. We employed the same real space meshing for both simulations at fundamental and second-harmonic frequencies, by checking that the second-harmonic one, corresponding to about 30 mesh steps per wavelength (i.e., to a minimum step mesh of 25 nm for both material platforms), was at convergence (i.e., resonance frequencies stable to within percent accuracy). The cavity modes are excited by point-like sources of electromagnetic radiation, once the emission of the sources is elapsed, the decaying fields are recorded in space and time domains. Such information allows to fully recover the frequencies and quality factors of both FH and SH cavity modes. The outcomes are reported in table~\ref{tab:cav}.

Moreover, from the FDTD simulations it is possible to retrieve the electromagnetic field profiles inside the cavities, as well as the emission properties of the cavity modes calculated from a near-to-far field projection of the 2D cavity modes recorded at the surface of the PhC slab. From the intensity profiles of the FH and SH modes that are shown  in Fig.~\ref{fig:cav_both}, it is clear how the radiation is well confined in the core region in both cases. The FH farfield intensity patterns are highly focused around the vertical direction of emission, which is crucial to facilitate the vertical in-coupling of the electromagnetic radiation coming from a pump beam with a gaussian intensity profile; while the SH mode exhibits a donut-shaped farfield intensity pattern, which can be directly attributed to its quasi-BIC nature \cite{art:momexp}.

\begin{figure*}
\centering
\includegraphics[width=0.95\linewidth]{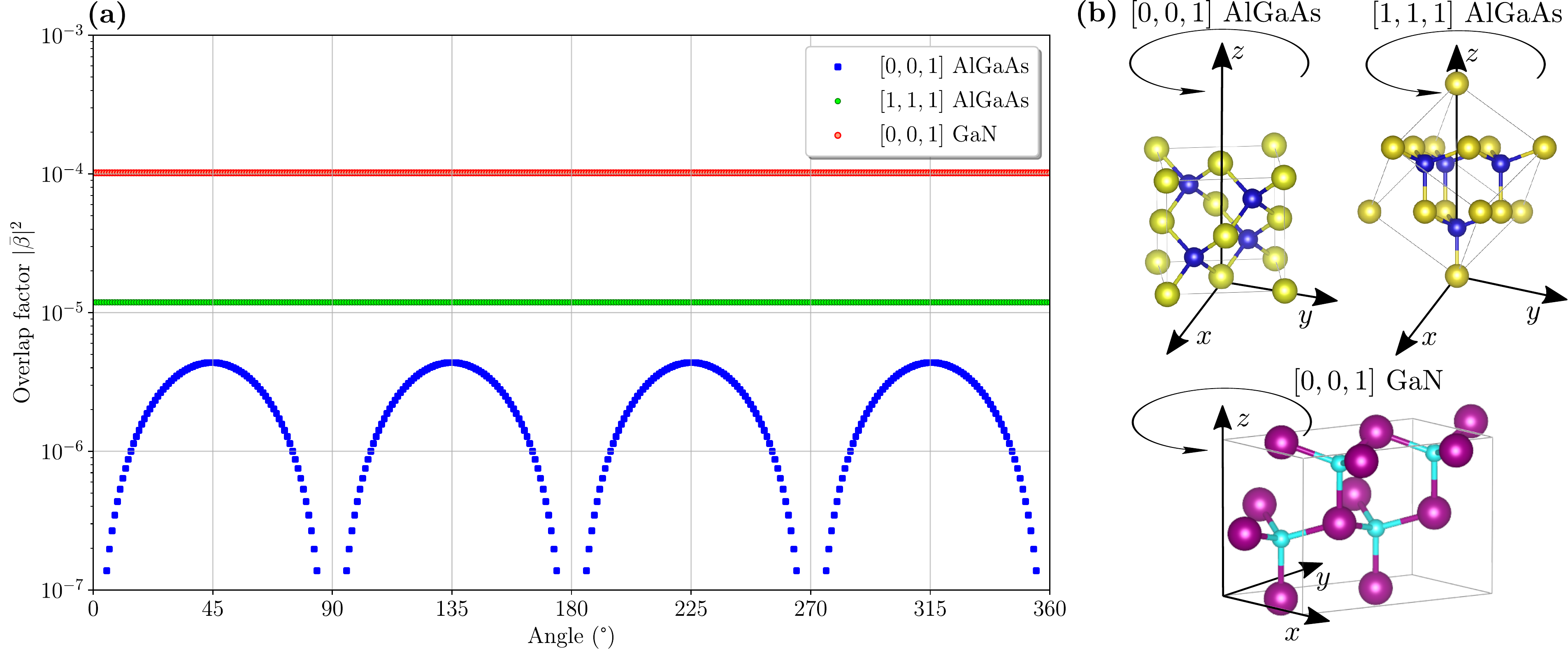}
\caption{\textbf{(a)} Nonlinear overlap factor between FH and SH modes in the doubly resonant PhC cavity design, $|\bar{\beta}|^2$, for \ch{GaN} and \ch{Al_{0.3}Ga_{0.7}As} grown along different directions, as a function of the rotation angle in the PhC lattice plane where \ang{0} indicates the [1,0,0] direction. A schematic of the unit cells and the definition of the rotation angles for the three cases considered is reported in \textbf{(b)}; these crystallographic pictures were obtained through the VESTA software \cite{art:crystals}. }\label{fig:overlap}
\end{figure*}   

\section{Overlap factor and nonlinear efficiency }\label{sec:overlap}
Having both spatial and temporal confinements of FH and SH modes is not a sufficient condition to guarantee an efficient second-order nonlinear efficiency. In fact, the spatial overlap between the FH and SH fields has to be suitably considered, as well as their confinement volume. 
For second-order nonlinear processes, a nonlinear overlap between the FH and SH modes can generally be defined as \cite{art:rodr}:
\begin{equation}\label{eq:beta}
    \bar{\beta}=\frac{\int d\mathbf{r}\,\sum_{ijk} \bar{\chi}^{(2)}_{ijk}E_{2\omega i}^*E_{\omega j}E_{\omega k} }{\left(\int d\mathbf{r}\,\varepsilon_{\omega}(\mathbf{r})|\mathbf{E}_\omega|^2 \right)\left(\int d\mathbf{r}\,\varepsilon_{2\omega}(\mathbf{r})|\mathbf{E}_{2\omega}|^2 \right)^{1/2}  }\sqrt{\lambda_\text{FH}^3},
\end{equation}
where $\mathbf{E}_\omega$ and $\mathbf{E}_{2\omega}$ are the FH and SH electric field profiles, $\varepsilon_{\omega}$ and $\varepsilon_{2\omega}$ are the dielectric functions at FH and SH frequencies, while $\bar{\chi}^{(2)}_{ijk}$ are dimensionless nonlinear tensor elements defined through $ \bar{\chi}^{(2)}_{ijk}=\chi^{(2)}_{ijk}/\chi^{(2)}$ with $\chi^{(2)}$ chosen to be the main tensor element of the relative compound grown in the $[0,0,1]$ direction:
\begin{equation}
\begin{aligned}
&\text{Zincblende}:\qquad \chi^{(2)}=\chi^{(2)}_{zxy} \\
&\text{Wurtzite}:\qquad\quad \chi^{(2)}=\chi^{(2)}_{xxz}.
\end{aligned}
\end{equation} 
The overlap factor in Eq.~\eqref{eq:beta} is expressed in dimensionless units, and it is particularly useful to compare different doubly resonant structures, also made of different materials. In fact, the constitutive material of the cavity and its crystallographic orientation are directly involved in the calculation of the integral in Eq.~\eqref{eq:beta} through the tensor elements $\bar{\chi}^{(2)}_{ijk}$.
Following the Kleinmann symmetry \cite{book:boyd}, the non-vanishing tensor elements of the  wurtzite (e.g., \ch{GaN}) and zincblende (e.g., \ch{Al_{0.3}Ga_{0.7}As}) crystal structures grown along the $[0,0,1]$ direction are: 
\begin{equation}\label{eq:tensor}
\begin{aligned}
&\text{Zincblende: }{\scriptstyle \bar{\chi}^{(2)}_{xyz}=\bar{\chi}^{(2)}_{zxy}=\bar{\chi}^{(2)}_{yzx}=\bar{\chi}^{(2)}_{xzy}=\bar{\chi}^{(2)}_{yxz}=\bar{\chi}^{(2)}_{zyx}=1 }\\
&\text{Wurtzite: }{\scriptstyle\, \bar{\chi}^{(2)}_{xxz}=\bar{\chi}^{(2)}_{zxx}=\bar{\chi}^{(2)}_{xzx}=\bar{\chi}^{(2)}_{yyz}=\bar{\chi}^{(2)}_{zyy}=\bar{\chi}^{(2)}_{yzy}=1, \; \bar{\chi}^{(2)}_{zzz}=-2.}
\end{aligned}
\end{equation}  
We supposed $\bar{\chi}^{(2)}_{zzz}=-2$, this condition is not always experimentally verified but since the FH is mostly TE-polarized, the actual value of the $\bar{\chi}^{(2)}_{zzz}$ tensor element has a negligible effect on the beta-factor of Eq.~\eqref{eq:beta}, as we have verified numerically.
Starting from \eqref{eq:tensor}, we can recover the tensor elements for an arbitrary orientation of the crystal axes with respect to the PhC lattice. This can be accomplished by applying a proper rotation in the three-dimensional space, described by a rotation matrix $\mathbf{R}(\phi,\theta,\psi)$ in which $(\phi,\theta,\psi)$ are the Euler angles (see App.~\ref{sec:rot} for details). For example, the transformation that allows to switch from the $[0,0,1]$ direction to the $[1,1,1]$ growth direction is
\begin{equation}
\begin{aligned}
&\bar{\chi}^{(2)\,[1,1,1]}_{\alpha,\beta,\gamma}=\sum_{a,b,c} U_{\alpha,a}U_{\beta,b}U_{\gamma,c}\, \bar{\chi}^{(2)\,[0,0,1]}_{a,b,c},  \\
& \mathbf{U}\doteq\mathbf{R}\left(\frac{3}{4}\pi,\cos^{-1}{\left(\frac{1}{\sqrt{3}}\right) },\frac{\pi}{2} \right).
\end{aligned}
\end{equation}

Here, we calculated the overlap integral \eqref{eq:beta} in the case of \ch{Al_{0.3}Ga_{0.7}As} grown both along the $[0,0,1]$ and $[1,1,1]$ directions. The \ch{GaN} is only considered to be grown along the $[0,0,1]$. Then, we also considered its dependence as a function of a rotation in the PhC lattice plane, i.e. the relative orientation between the PhC lattice and the crystallographic directions of the underlying nonlinear material. The squared modulus of the overlap integral is reported in Fig.~\ref{fig:overlap}(a) for both material platforms. A definition of the rotation angle is schematically given in Fig.~\ref{fig:overlap}(b).

\begin{figure}[b]
\centering
\includegraphics[width=\linewidth]{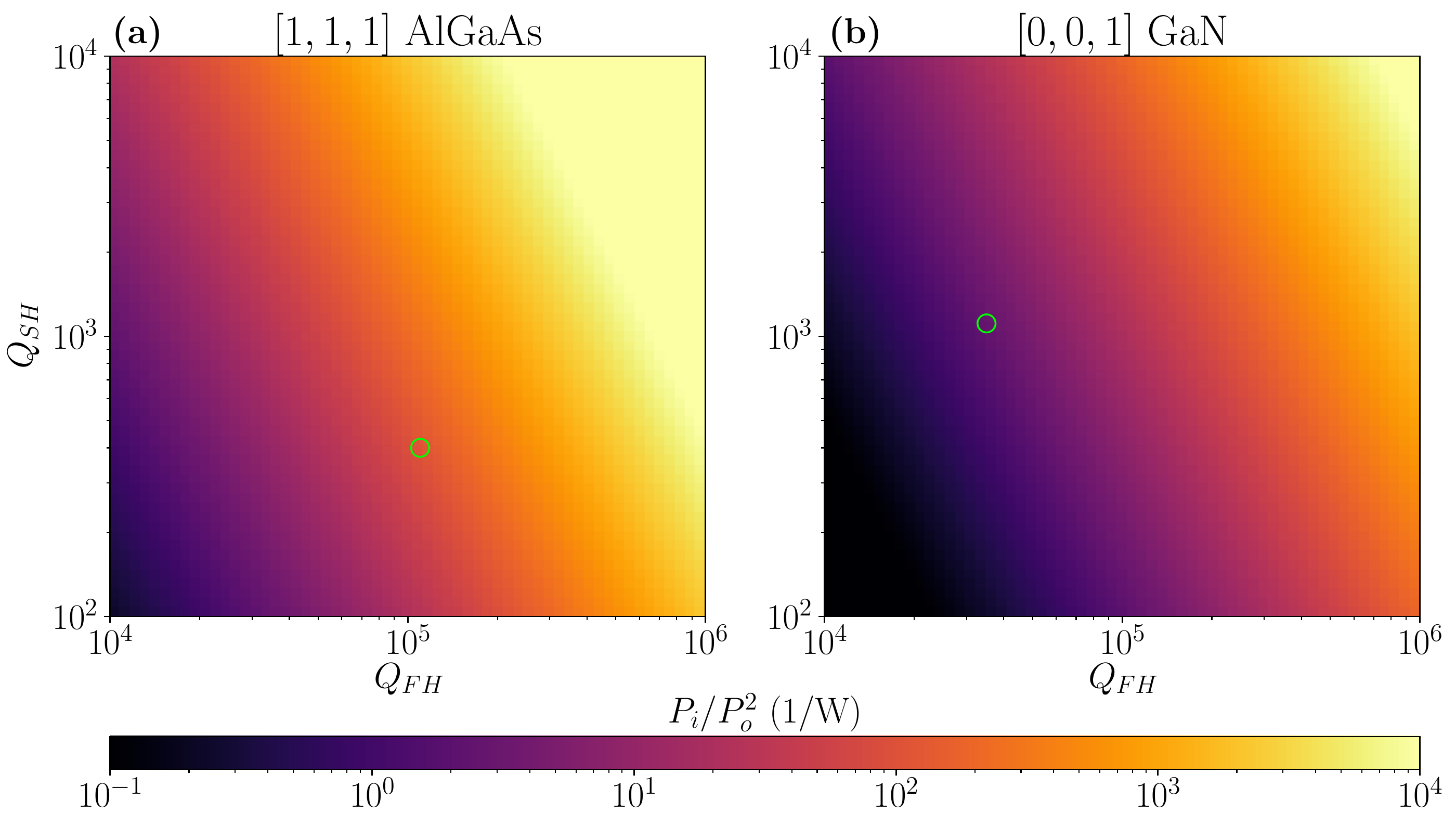}
\caption{\textbf{(a)}-\textbf{(b)} SHG conversion efficiency for the \ch{Al_{0.3}Ga_{0.7}As} and \ch{GaN} doubly resonant PhC cavities, calculated for varying FH and SH cavity factors. The green circles identify the expected efficiency for the cavities designed in the present work.}\label{fig:eff}
\end{figure}

\begin{table*}
\caption{Calculated nonlinear figures of merit for both the \ch{GaN} and \ch{Al_{0.3}Ga_{0.7}As} doubly resonant PhC cavities designed in this work.}\label{tab:cav_res}
\centering
\begin{tabular}{cccccc}
\toprule
\textbf{Material}  &$Q_{\text{FH}}$ &$Q_{\text{SH}}$	& $|\bar{\beta}_{\text{max}}|^2$&$\chi^{(2)}$ (\si{\pico\meter\per\volt})	& $P_o/P_i^2$ (\si{\per\watt})  \\
\ch{GaN}		& $3.6\cdot10^4$  &$1100$ 		& $ 1.0\cdot10^{-4}$& $10$  & $2.9$\\
\ch{Al_{0.3}Ga_{0.7}As}	 & $1.1\cdot10^5$	& $400$		& $ 1.2\cdot10^{-5}$	& $100$ &$119$ \\
\toprule
\end{tabular}
\end{table*}

The overlap factor, $|\bar{\beta}|^2$, displays different qualitative trends depending on the crystal structure and its relative orientation. In particular, the largest nonlinear overlap is obtained for the \ch{GaN} where $|\bar{\beta}|^2$ goes beyond $10^{-4}$ for any orientation angle, while in the \ch{Al_{0.3}Ga_{0.7}As} it is modulated and can vanish at specific orientation angles for the $[0,0,1]$ growth direction. On the other hand, the $|\bar{\beta}|^2$ does not depend on the relative orientation between the crystallographic directions and the PhC lattice for the \ch{Al_{0.3}Ga_{0.7}As} grown along the $[1,1,1]$, which also displays an overall larger values as compared to the same material grown along $[0,0,1]$.

While the dimensionless overlap factor gives the dependence of the conversion efficiency on the crystallographic axes orientation, it only gives partial information since the physical process of nonlinear conversion involves the actual magnitude of the $\chi^{(2)}$ tensor elements, which is material dependent and is not explicitly taken into account in Eq.~\ref{eq:beta}. This additional parameter can play a crucial role in calculating the final nonlinear efficiency, which can be estimated with the following relation \cite{art:rodr}:
\begin{equation}\label{eq:eff_fin}
    \frac{P_o}{P_i^2}=\frac{8}{\omega_1}\left(\frac{\chi^{(2)}} {\sqrt{\varepsilon_0\lambda^3_{\text{FH}}}}\right)^2|\bar{\beta}|^2\,
    Q^2_{\text{FH}}Q_{\text{SH}},
\end{equation}
where $\varepsilon_0$ is the vacuum permittivity. A detailed derivation of Eq.~\eqref{eq:eff_fin} is reported in App.~\ref{sec:eff} for completeness. We also notice that Eq.~\eqref{eq:eff_fin} is valid under the undepleted-pump approximation, and considering perfect in- and out-coupling of the radiation for both FH and SH modes \cite{art:momch}. Notice that similar scaling relations can be found for other nonlinear generation processes, e.g., SPDC \cite{art:Helt2012}. Notice also that this expression is explicitly obtained for narrow-band harmonic resonances, at variance with SHG in broadband photonic crystal waveguide structures  \cite{art:rivoireAPL2011}.
We evaluated the conversion efficiencies for a wide range of $Q_{\text{FH}}$ and $Q_{\text{SH}}$. We set $\chi^{(2)}=$ \SI{100}{\pico\meter\per\volt} for \ch{Al_{0.3}Ga_{0.7}As}, and  $\chi^{(2)}=$ \SI{10}{\pico\meter\per\volt} for \ch{GaN}, respectively \cite{art:nonl_AlGaAs,art:nonl_GaN}. Moreover, we considered the axes orientations maximizing the nonlinear overlap factor according to the results in Fig.~\ref{fig:overlap}(a). The results are reported in Fig.~\ref{fig:eff}, which inverts the outcome of the previous Figure: in fact, the \ch{Al_{0.3}Ga_{0.7}As} doubly resonant cavity (grown along [1,1,1]) is theoretically more efficient than the \ch{GaN} one, despite the dimensionless $|\bar{\beta}|^2$ factor being one order of magnitude larger in favor of the latter. This can mostly be attributed to the larger nonlinear properties of  \ch{Al_{0.3}Ga_{0.7}As}, which is characterised by an higher $\chi^{(2)}_{\text{eff}}$ .
If we restrict ourselves to the quality factors calculated in the present work with perfectly in- and out-coupled beams, we can estimate an efficiency of \SI{119}{\per\watt} for the \ch{Al_{0.3}Ga_{0.7}As}, and \SI{2.9}{\per\watt} for the \ch{GaN} doubly resonant PhC cavity, as summarized in table~\ref{tab:cav_res}.

\section{Conclusions}
We have presented a numerical procedure to design doubly-resonant PhC cavities by using evolutionary optimization algorithms, which can be applied to a vast selection of cavity geometries and material platforms. The main figures of merit, such as the quality factors of the cavity modes and their nonlinear field overlap, were optimized targeting the main III-V semiconductors employed in nonlinear photonics. These results could help the practical realization of such devices for applications in high-efficiency second-order nonlinear processes.


\acknowledgments{We acknowledge useful discussions with M. Bollani, M. Clementi, C. De Angelis, M. Galli, M. Liscidini. }
\section*{Funding}
This research was partly supported from the Italian Ministry of Education and Research (MIUR) through 2017 PRIN project 2017MP7F8F-004 ``NOMEN''. M. Minkov and S. Fan acknowledge the support of a MURI project from the U.S. Air Force Office of Scientific Research (Grant No. FA9550-17-1-0002).



\appendix

\section{Rotation of the nonlinear tensor }\label{sec:rot} 
\unskip
\subsection{Euler angles and rotation matrices}
\begin{figure*}
\centering
\includegraphics[width=0.7\linewidth]{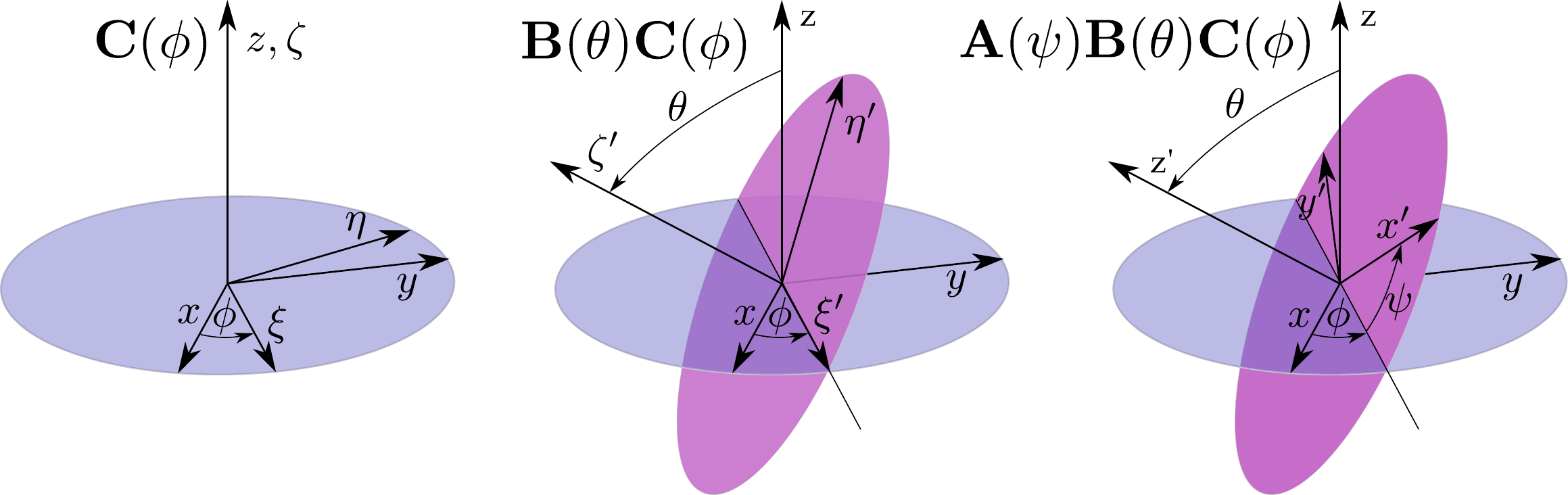}
\caption{Definition of the Euler angles describing an arbitrary 3D rotation, for which the reference frame $(x,y,z)$ is transformed into $(x',y',z')$ by applying the rotation matrix $\mathbf{R}(\phi,\theta,\psi)$.}\label{fig:euler_ang}
\end{figure*}

Every rotation in the 3D space can be expressed in terms of the Euler angles $(\phi,\theta,\psi)$, which define the rotation matrix $\mathbf{R}(\phi,\theta,\psi)= \mathbf{A}(\psi)\mathbf{B}(\theta)\mathbf{C}(\phi)$ \cite{book:goldstein}, where
\begin{align*}\scriptstyle
\mathbf{A}\left(\psi\right)= 
\begin{pmatrix}
\scriptstyle \cos\psi & \scriptstyle\sin\psi & \scriptstyle0 \\
\scriptstyle\scriptstyle-\sin\psi & \scriptstyle\cos\psi &\scriptstyle 0 \\
\scriptstyle0 &\scriptstyle 0 &\scriptstyle 1  
\end{pmatrix},\\ \scriptstyle
\mathbf{B}\left(\theta\right)=
\begin{pmatrix}
\scriptstyle1 &\scriptstyle 0 &\scriptstyle 0 \scriptstyle\scriptstyle\\
\scriptstyle0 & \scriptstyle\cos\theta & \scriptstyle\sin\theta \\
\scriptstyle0 &\scriptstyle -\sin\theta &\scriptstyle \cos\theta 
\end{pmatrix},\\\scriptstyle
\mathbf{C}\left(\phi\right)=
\begin{pmatrix}
\scriptstyle\cos\phi &\scriptstyle \sin\phi &\scriptstyle 0 \\
\scriptstyle-\sin\phi & \scriptstyle\cos\phi &\scriptstyle 0 \\
\scriptstyle0 &\scriptstyle 0 &\scriptstyle 1 
\end{pmatrix}. 
\end{align*}
The coordinate transformation $(x,y,z)\to(x',y',z')$ obtained as a result of the generic rotation $\mathbf{R}(\phi,\theta,\psi)$ is schematically illustrated in Fig.~\ref{fig:euler_ang}.

\subsection{Rotated nonlinear tensor}
The reduced nonlinear tensor, $\bar{d}$ (see Ref.~\cite{book:boyd} for a definition in terms of $\chi^{(2)}$ tensor elements), is hereby reported for the zincblende (wurtzite) in the reference frame corresponding to the $[0,0,1]_z$ and $[1,1,1]_z$ ($[0,0,1]_w$) crystal growth directions. 

\begin{align*}
2\bar{d}=\begin{cases}
[0,0,1]_z:&
\begin{pmatrix}
\scriptstyle0 &\scriptstyle 0 &\scriptstyle 0 &\scriptstyle1&\scriptstyle0&\scriptstyle0 \\
\scriptstyle0 &\scriptstyle 0 &\scriptstyle 0 &\scriptstyle0&\scriptstyle1&\scriptstyle0 \\
\scriptstyle0 &\scriptstyle 0 &\scriptstyle 0 &\scriptstyle0&\scriptstyle0&\scriptstyle1 
\end{pmatrix}, \\\\[7pt]
[1,1,1]_z:&
\begin{pmatrix}
\scriptstyle\sqrt{\frac{2}{3} } &\scriptstyle -\sqrt{\frac{2}{3}}  &\scriptstyle 0 &\scriptstyle 0 &\scriptstyle -\sqrt{\frac{1}{3}}&\scriptstyle0 \\
\scriptstyle0 &\scriptstyle 0 &\scriptstyle 0 &\scriptstyle   -\sqrt{\frac{1}{3}}  &\scriptstyle 0 &\scriptstyle -\sqrt{\frac{2}{3}}  \\
\scriptstyle-\sqrt{\frac{1}{3}} &\scriptstyle -\sqrt{\frac{1}{3}} &\scriptstyle \sqrt{\frac{4}{3} } &\scriptstyle0&\scriptstyle0&\scriptstyle0 
\end{pmatrix},\\
\end{cases}
\end{align*}
\begin{align*}
2\bar{d}=\begin{cases}
   [0,0,1]_w: &
\begin{pmatrix}
\scriptstyle0 &\scriptstyle 0 &\scriptstyle 0 &\scriptstyle0&\scriptstyle1&\scriptstyle0 \\
\scriptstyle0 &\scriptstyle 0 &\scriptstyle 0 &\scriptstyle1&\scriptstyle0&\scriptstyle0 \\
\scriptstyle1 &\scriptstyle 1 &\scriptstyle -2 &\scriptstyle0&\scriptstyle0&\scriptstyle0 
\end{pmatrix}.\qquad
\end{cases}
\end{align*}

\section{Particle swarm optimization}\label{sec:PSO}
The particle swarm optimization (PSO) is an evolutionary algorithm employed for the search of the minimum (or the maximum) of a given function $f:A\to\mathbb{R}$ in the search space $A\subset\mathbb{R}^n$. Firstly, the PSO generates a certain number, D, of trial solutions $\mathbf{x}\in A$, these define the D-dimensional first generation. The subsequent generations are calculated by evolving the previous solutions with the formula $\mathbf{x}_t=\mathbf{x}_{t-1}+\mathbf{v}_t$, where $t$ labels the generation, while $\mathbf{v}_t$ is evaluated as follows:
\begin{align*}
    \mathbf{v}_t=\omega\mathbf{v}_{t-1}+c_1\eta_1(\mathbf{p}_{t-1}-\mathbf{x}_{t-1})+c_2\eta_2(\mathbf{g}_{t-1}-\mathbf{x}_{t-1}),
\end{align*}
where $\omega$ is an inertial weight, $\eta_{1,2}$ are two random numbers between $0$ and $1$, $\mathbf{p}$ is the best position explored by $\mathbf{x}$, $\mathbf{g}$ is the global best position, while $c_1$ and $c_2$ are the cognitive and social factors, respectively.

In the present work, the Lumerical-Ansys implementation of the PSO has been used, in which standard values of $\omega$, $c_1$ and $c_2$ guarantee the fast algorithm convergence in most cases.

\section{Second-harmonic conversion efficiency}\label{sec:eff}
The temporal coupled rate equations of two cavity modes, $a_1$ and $a_2$, which mutually exchange energy through the coupling terms $\beta_1$ and $\beta_2$ can be formally written as \cite {art:rodr}:
\begin{align}
&\frac{da_1}{dt}=i\omega_1\left(1 +\frac{i}{2Q_1} \right)a_1-i\omega_1\beta_1a_1^*a_2+\sqrt{\frac{2}{\tau_{1}} }s_{1+},\label{eq:r_1}\\
&\frac{da_2}{dt}=i\omega_1\left(1 +\frac{i}{2Q_2} \right)a_2-i\omega_2\beta_2a_1^2,\label{eq:r_2}\\
&s_{1-}=\sqrt{\frac{\omega_1}{Q_1}}a_1-s_{1+}\label{eq:r_s1},\\
&s_{2-}=\sqrt{\frac{\omega_2}{Q_2}}a_2\label{eq:r_s_2},
\end{align}
where $Q_i=\omega_i\tau_i/2$ ($i=1,2$) are the quality factors of the cavity modes, $s_{1+}$ is the pump term for the first mode, while $s_{i-}$ describe the output rates. By assuming oscillating solutions, we set $a_1=\tilde{a}_1e^{i\omega t}$, $a_2=\tilde{a}_2e^{i2\omega t}$, $s_1=\tilde{s}_{1+}e^{i\omega t}$, and neglecting the down-conversion term $i\omega_1\beta_1a_1^*a_2$ from Eq. \ref{eq:r_1}, we get
\begin{equation}\label{eq:r_1_sol}
\frac{d\left(\tilde{a}_1 e^{i\omega t}\right)}{dt}=i\omega_1\left(1 +\frac{i}{2Q_1} \right)\tilde{a}_1e^{i\omega t}+\sqrt{\frac{2}{\tau_{1}} }\tilde{s}_{1+}e^{i\omega t} \, .
\end{equation}
In steady state the condition $d\tilde{a}_1/dt=0$ applies, hence Eq.~\ref{eq:r_1_sol} simplifies to
\begin{equation}
i\omega\tilde{a}_1=i\omega_1\tilde{a}_1-\frac{1}{\tau_1}\tilde{a}_1+\sqrt{\frac{2}{\tau_{1}} }\tilde{s}_{1+} \, ,
\end{equation}
leading to the following relation between $|\tilde{a}_1|^2$ and $|\tilde{s}_{1+}|^2$
\begin{equation}\label{eq:a_s_1}
|\tilde{a}_1|^2=\frac{2/\tau_1}{(\omega-\omega_1)^2+1/\tau_1^2}\,|\tilde{s}_{1+}|^2 \, .
\end{equation}
Following the same procedure with Eq.~\ref{eq:r_2}, we obtain 
\begin{equation}
\frac{d\left(\tilde{a}_2 e^{i2\omega t}\right)}{dt}=i\omega_2\left(1 +\frac{i}{2Q_2} \right)\tilde{a}_2e^{i2\omega t}-i\omega_2\beta_2\left(\tilde{a}_1e^{i\omega t}\right)^2 \, .
\end{equation}
As in the previous case, steady state solutions can be found by imposing the condition $d\tilde{a}_2/dt=0$, thus obtaining
\begin{equation}
i2\omega\tilde{a}_2=i\omega_2\tilde{a}_2-\frac{1}{\tau_2}\tilde{a}_2-i\omega_2\beta_2\tilde{a}_1^2 \, ,
\end{equation}
which leads to the expression that links $|\tilde{a}_2|^2$ with $|\tilde{a}_1|^4$:
\begin{equation}\label{eq:a_s_2}
|\tilde{a}_2|^2=\frac{\omega_2^2\,|\beta_2|^2}{(2\omega-\omega_2)^2+1/\tau_2^2 }\, |\tilde{a}_1|^4 \, .
\end{equation}
Finally, we can calculate the conversion efficiency as $P_o/P_i^2=|\tilde{s}_{2-}|^2/|\tilde{s}_{1+}|^4$, in which $|\tilde{s}_{2-}|^2$ is straightforwardly calculated from Eq. \ref{eq:r_s_2} as
\begin{equation}\label{eq:s_2}
|\tilde{s}_{2-}|^2=\frac{\omega_2}{Q_2}|\tilde{a}_2|^2=\frac{2}{\tau_2}|\tilde{a}_2|^2 \, ,
\end{equation}
in which we can insert Eqs.~\ref{eq:a_s_1} and \ref{eq:a_s_2} after imposing the resonant conditions, i.e., $2\omega=\omega_2$ and $\omega=\omega_1$, from which we get
\begin{equation}\label{eq:s_1_2}
|\tilde{s}_{2-}|^2=8\omega_2^2|\beta_2|^2\tau_2\tau_1^2||\tilde{s}_{1+}|^4 = \frac{128}{\omega_1}|\beta_2|^2Q_1^2Q_2|\tilde{s}_{1+}|^4 \, .
\end{equation}
Finally, given the expression of $\beta_2$
\begin{equation}
\beta_2 =\frac{1}{4}\frac{\int d\mathbf{r}\; \chi^{(2)}\bar{\chi}_{ijk}E_{2\omega i}^*E_{\omega j}E_{\omega k} }{\left(\int d\mathbf{r}\, \varepsilon_0 \varepsilon_1 |\mathbf{E}_\omega|^2\right)\left(\int d\mathbf{r} \,\varepsilon_0 \varepsilon_2 |\mathbf{E}_{2\omega}|^2\right)^{\frac{1}{2}}} \, ,
\end{equation}
a dimensionless $\bar{\beta}$ can be defined as
\begin{equation}
\bar{\beta} =\frac{\int d\mathbf{r}\; \bar{\chi}_{ijk}E_{2\omega i}^*E_{\omega j}E_{\omega k} }{\left(\int d\mathbf{r}\,  \varepsilon_1 |\mathbf{E}_\omega|^2\right)\left(\int d\mathbf{r} \, \varepsilon_2 |\mathbf{E}_{2\omega}|^2\right)^{\frac{1}{2}}} \, ,
\end{equation}
where $\varepsilon_1=\varepsilon(\omega_1)$ and $\varepsilon_2=\varepsilon(\omega_2)$. 
The conversion efficiency can then be obtained from Eq. \ref{eq:s_1_2}, in terms of $\bar{\beta}$, $Q_1$ and $Q_2$, as
\begin{equation}
\frac{P_o}{P_i^2}=\frac{8}{\omega_1}\left(\frac{\chi^{(2)}}{\sqrt{\varepsilon_0\lambda_1^3} } \right)^2|\bar{\beta}|^2Q_1^2Q_2 \, ,
\end{equation}
which reproduces Eq.~\eqref{eq:eff_fin} in the main text, and it is consistent with the expression already reported in the literature \cite{art:rodr}.

\bibliographystyle{apsrev4-1}
\bibliography{Doubly}

\end{document}